# Flat-top interleavers based on single MMIs

Matteo Cherchi*, Fei Sun, Markku Kapulainen, Mikko Harjanne, and Timo Aalto
VTT – Technical Research of Finland, Tietotie 3, 02150 Espoo, Finland

## ABSTRACT

We demonstrate for the first time flat-top interleavers based on cascaded Mach-Zehnder interferometers (MZIs) which use only single multimode interferometers (MMIs) as power splitters. Our previous designs were based on 4-stage cascades of MZIs, where we used single MMIs and double MMIs to achieve 85:15 splitting ratio and 31:69 splitting ratio respectively. This time, we propose instead a greatly simplified 2-stage configuration using only single MMIs, including a standard 50:50 MMI, and two tapered MMIs to achieve 71:29 and 92:08 splitting ratios. We have designed the interleaver based on its geometrical representation on the Bloch sphere, then confirmed by efficient 2D simulations of the building blocks and of the whole structure, based on the eigenmode expansion method. We show how important is to take into account the phase relations between the outputs of all MMIs in order to make a working design. We have successfully fabricated devices with different channel spacing on our micron-scale silicon photonics platform, and measurement results confirmed their expected flat-top operation on a broad band. Using only single MMI splitters we can not only greatly outperform the bandwidth achieved by standard directional couplers, but we can also ensure much higher robustness to fabrication errors, also compared to previous demonstrations based on double MMIs. Indeed, when compared to those previous attempts, the new results prove tapered MMIs to be the most robust approach to achieve arbitrary splitting ratios.

## 1. INTRODUCTION

Interleavers are enabling basic building blocks for optical communication that help, for example, multiplexing and demultiplexing closely spaced narrow channels in wavelength division multiplexing (WDM) systems[1]. By separating adjacent channels in different physical paths, interleavers help relaxing the requirements for multiplexers in terms of cross-talk from adjacent channels, therefore enhancing the bandwidth utilization of each channel. The simplest realization of an interleaver is a simple unbalanced MZI, with suitable unbalance between the two arms. Nevertheless, the sinusoidal response of such a device is far from the ideal box-like shape desirable for band-pass and band-stop filters. For this reason, several solutions have been proposed to achieve flat-top response. We list below the most relevant ones.

A first approach exploits the $\pi$-jump across the resonance of a ring resonator placed on the shorter arm of a MZI[2–4]. This way, interleavers with very steep roll-off can be realized, even though a trade-off with in-band ripple and extinction ratio must be found. In a recent publication, we have demonstrated that very robust ring-loaded MZIs can be achieved by using MMI splitters to couple the ring to the MZI arm[5]. A limitation of this approach is that ring resonators cannot have arbitrary large free spectral range (FSR).

A second approach exploits symmetries in so called lattice filters[4] (also known as Fourier filters) with same imbalance in all stages, and in particular in doubly-symmetric 4-stage MZIs[6]. The symmetries ensure auto-compensation of the phase errors that occur when the wavelength departs from the central wavelength of a given channel, i.e. when the relative phase between the two arms departs from multiples of $\pi$. It was later demonstrated by one of us[7] that the class of suitable symmetries and working devices can be extended, by using a simple analytic approach, enabled by a powerful geometrical representation. This provided much deeper physical insight compared to the numerical optimization approach of the original paper. Recently, we have demonstrated how such 4-stage flat-top filters can be realized using MMI splitters in all stages[8]. The main issue with this approach is the limited adjustability of in-band ripple, roll-off and extinction ratio.

A somewhat similar approach is based on generalized lattice filters[4,9] where the imbalance between the two arms doubles at each stage. Also in this case, the usual design approach is based on numerical optimization algorithms to determine the coupling coefficients of the splitters. Compared to standard lattice filters, generalized lattice filters can achieve similar performance with fewer stages. For example, a 2-stage generalized lattice filter can achieve comparable flat-top band and roll-off of a standard 4-stage lattice filter.

*matteo.cherchi@vtt.fi; https://www.vtt.fi/sites/siliconphotonics

In this paper, we first show how 2-stage generalized lattice filters can be interpreted and designed based on a geometrical representation and analytical formulas derived from the same. We then show how to implement the design on a micron-scale silicon photonic platform, by using only single MMIs, which splitting ratios can be continuously changed by simply tapering up or down canonical MMIs. Eventually, we report the results of optical characterisation of fabricated MMIs and interleavers.

# 2. GEOMETRIC INTERPRETATION

## 2.1 Geometric representation

Interferometric 2-waveguide systems are combinations of two basic building blocks, namely power splitters – like directional couplers or MMI splitters – and phase shifters – like length imbalances or waveguide tapering[10]. They can be treated analytically based on transfer matrix, but significant physical insight can be gained when representing them on the so-called Bloch sphere[10–12] (also known as Poincaré sphere, when used for polarisation states). All possible combinations of power splitting ratio and relative phase can be mapped on the sphere as depicted in Figure 1. Therefore, any state is unequivocally determined by two angles: $\theta$ representing the phase shift and $2\alpha$ corresponding to the splitting ratio $\sin^2(\alpha)/\cos^2(\alpha)$.

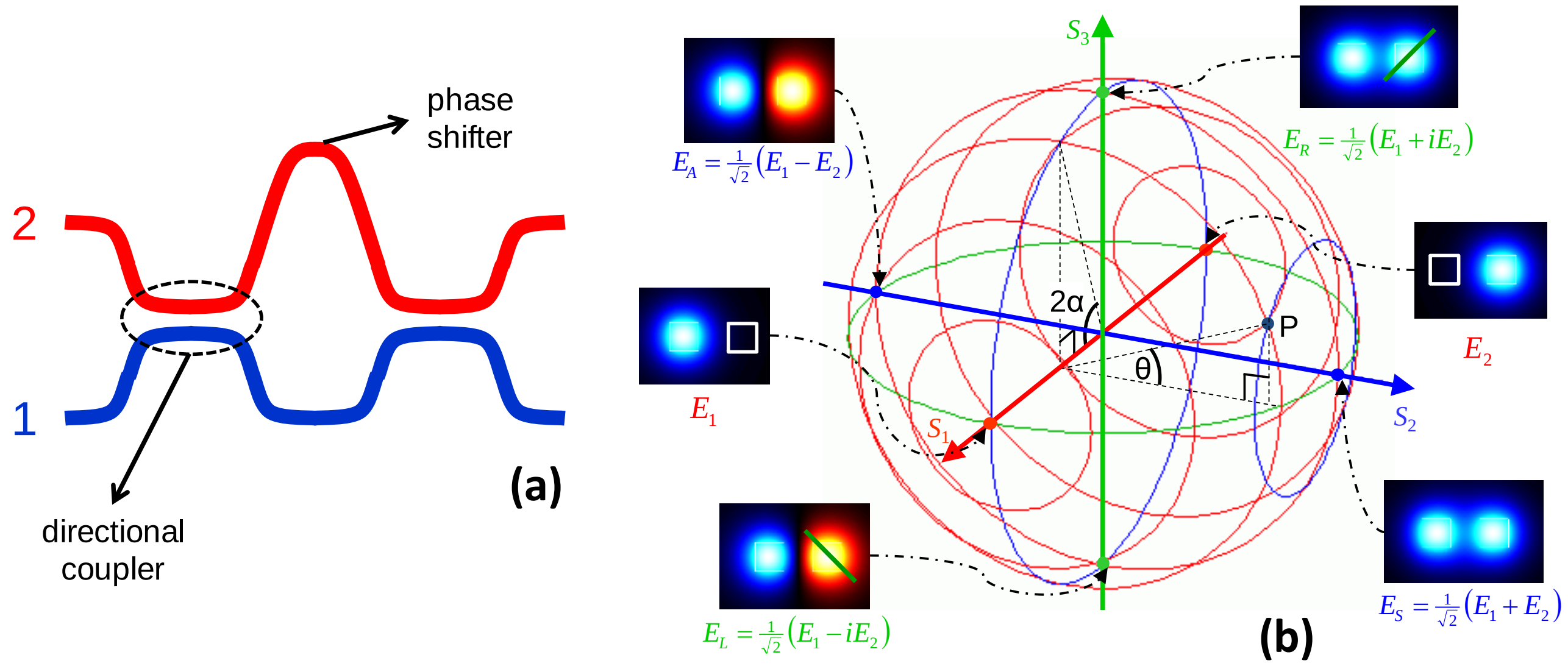

Figure 1. a) Schematic layout of a simple 2-waveguide interferometer; b) the Bloch sphere can be used as a geometric representation of all possible states of a 2-waveguide system.

The three Stokes parameters $S_1$, $S_2$, and $S_3$ correspond to the three orthogonal axis. The intersections of $S_1$ with the sphere represent the single uncoupled waveguide modes $E_1$ and $E_2$, the intersections of $S_2$ represent the eigenmodes of synchronous directional couplers, i.e. the symmetric and anti-symmetric modes $E_S$ and $E_A$, whereas the intersections of $S_3$ represent the superpositions in quadrature and anti-quadrature $E_R$ and $E_L$.

Remarkably, the operation of splitters and phase shifters can be represented as simple rotations on the sphere. A synchronous directional coupler is a rotation around $S_2$, which is the axis of its eigenmodes, i.e. the symmetric and anti-symmetric modes $E_S$ and $E_A$. Similarly, a phase shifter is a rotation around $S_1$, which is the axis of its eigenmodes, i.e. the single waveguide modes $E_1$ and $E_2$. As an example, we show in Figure 2 the operation of a MZI with two 50:50 directional couplers and a $\pi/4$ phase shift in between. The splitting ratio at the output is simply determined by the angle $\alpha$. The figure intuitively show also an important property of MZIs: the final point always ends up on the equator, meaning that the outputs are always either perfectly in phase or completely out of phase.

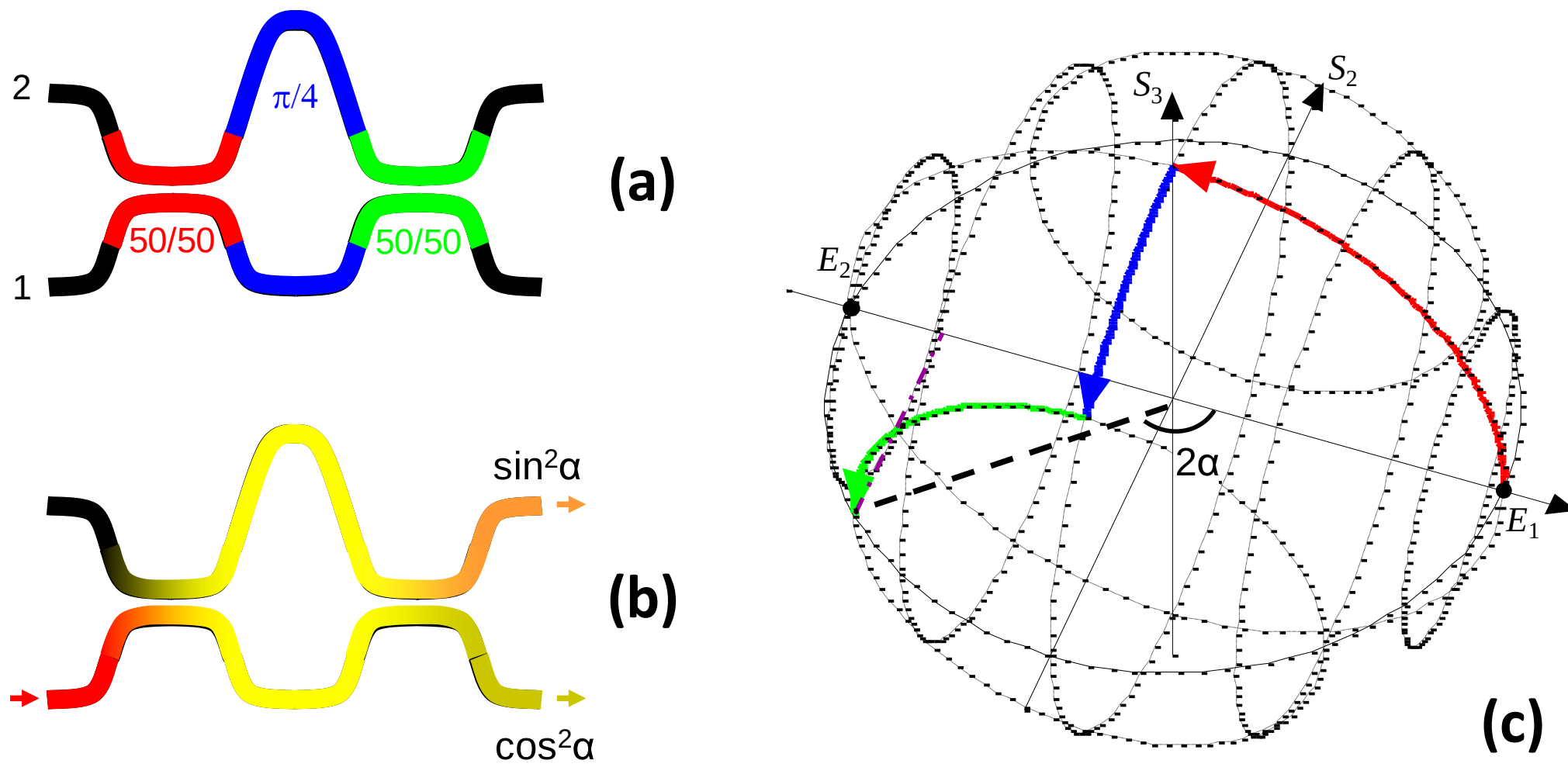

Figure 2. a) Schematic of a MZI with π/4 phase shifter; b) resulting power splitting in terms of the angle α; c) geometric representation of the MZI operation.

## 2.2 Interpretation of 2-stage flat-top filters

The geometric representation helped us understanding the working principle of 4-stage doubly-symmetric lattice filters, showing how the flat-top response is achieved thanks to the auto-corrective behaviour of suitable phase shifter combinations[7,8]. That physical insight enabled us to extend the class of working devices, and to derive their splitting ratios as simple analytical formulas, instead of using the numerical optimisation algorithms adopted in previous approaches.

We show here how similar insight can be gained also for 2-stage generalised lattice filters. We start from one of the examples reported by Madsen and Zhao[4] (page 221 of their book, Figure 4-49,), with couplers of 0.500, 0.7143, and 0.9226, and phase shifters of order +1 and +2 (but the interpretation can be easily extended to any of the other examples reported). We sketch in Figure 3a a possible physical realisation of the filter, with the phase shifters implemented as arms of different physical length. The transfer matrix response of the filter is plotted in Figure 3c. Central wavelengths of transmission bands in the cross-port correspond to a $\pi+2k\pi$ ($k$ integer) phase shift in the first phase shifter, whereas for the bar port, central wavelengths correspond to $2m\pi$ ($m$ integer) phase shift in the first phase shifter. For wavelengths slightly off the central wavelengths, the phase shift will depart from multiples of $\pi$, which in the case of simple MZIs leads to reduced transmission and extinction ratio, i.e. to non-flat response. Instead, the 2-stage filter is designed such that the phase offset in the first stage is compensated by the offset in the second stage, resulting in autocorrective behaviour and consequent flat response. In fact, the angular offset in the second stage is, by construction twice as the angular offset in the first stage, but it occurs on a circle that has about half the radius, and in the opposite direction. This is shown in Figure 3b and c, for the cross port and bar port respectively. The half radius is guaranteed by having chosen a 71:29 splitter, i.e. very close to 75:25 that would correspond to exactly 120° rotation. This way, the physical offsets cancel out, having same magnitude and opposite sign. The working principle somewhat resembles the one of flat-top interferometric splitters with 120° phase shifters[10,13]. Choosing a rotation slightly short of 120° helps ensuring even a wider flat-top operation, and can be fine-tuned to find a trade-off with roll-off, extinction ratio, and in-band ripple.

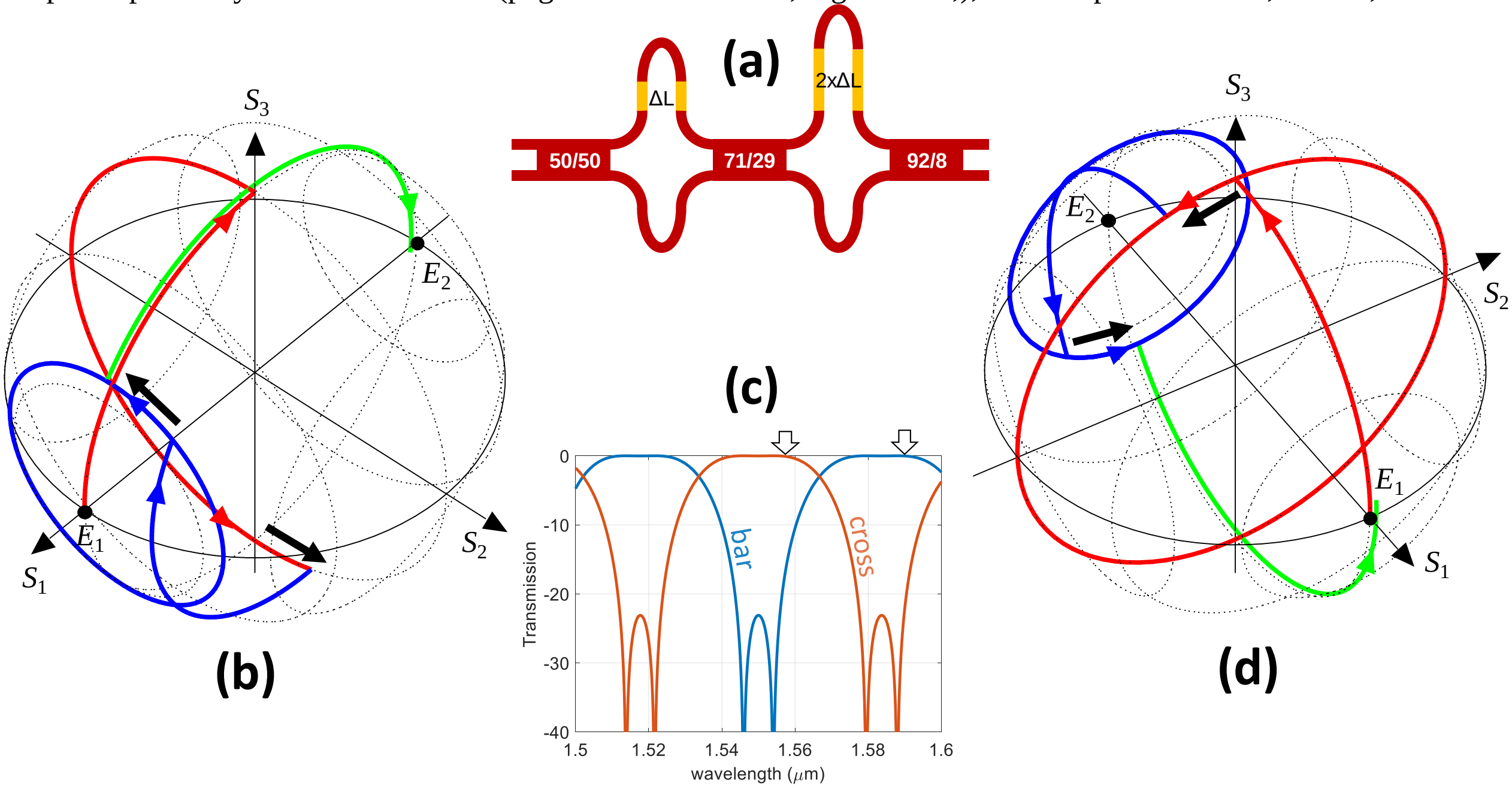

Figure 3. a) Schematic layout of the 2-stage flat-top generalised lattice filter; b) Operation of the filter for a wavelength slightly off the centre of the cross channel, as shown by the arrow in c). d) The same for a wavelength slightly off the centre of the bar channel. The black arrows in b) and d) highlight the auto-corrective behaviour of the filter.

### 2.3 Analytical formulas

Thanks to the geometric representation, we can now write analytic expression to identify all possible working configurations. If, like in the previous example, we want $\pi+2k\pi$ phase shifts and $2m\pi$ phase shifts to correspond to the cross state and to the bar state respectively, the rotation angles on the sphere $\phi_1$, $\phi_2$, and $\phi_3$ corresponding to the three splitters must obey the equation system

$$\begin{cases} \phi_1 + \phi_2 + \phi_3 = 2m\pi \\ \phi_2 - \phi_1 + \phi_3 = (2k+1)\pi \\ |\sin\phi_1| = 2|\sin(\phi_2 - \phi_1)| \end{cases} \tag{1}$$

Where the last relation is to impose that the radius of the second phase rotation is half the radius of the first phase rotation. We can determine $\phi_1$ by subtracting the second equation from the first equation, and then use that result to determine $\phi_2$ from last equation. Eventually $\phi_3$ can be also derived, resulting in the following solutions:

$$\begin{cases} \phi_1 = \frac{\pi}{2} + p\pi \\ \phi_2 = \frac{-3\pm1}{6}\pi + q\pi \\ \phi_3 = \mp\frac{\pi}{6} + t\pi \end{cases}, \tag{2}$$

where $p$, $q$ and $t$ are integers. In particular, the solution $\phi_1 = \pi/2$, $\phi_2 = 2\pi/3$, and $\phi_3 = 5\pi/6$ corresponds to the filter in Figure 3. Some of the solutions, including this one, require same sign of the two phase shifters, whereas some other solutions require phase shifters with opposite sign.

Similar procedure can be followed to identify the solutions when $\pi+2k\pi$ phase shifts and $2m\pi$ phase shifts correspond instead to the bar state and to the cross state respectively:

$$\begin{cases} \phi_1 + \phi_2 + \phi_3 = (2k+1)\pi \\ \phi_2 - \phi_1 + \phi_3 = 2m\pi \\ |\sin\phi_1| = 2|\sin(\phi_2 - \phi_1)| \end{cases} \Rightarrow \begin{cases} \phi_1 = -\frac{\pi}{2} + p\pi \\ \phi_2 = \frac{3\pm1}{6}\pi + q\pi. \\ \phi_3 = \mp\frac{\pi}{6} + t\pi \end{cases} \tag{3}$$

For example, the solution $\phi_1 = \pi/2$, $\phi_2 = \pi/3$, and $\phi_3 = \pi/6$ (figure 4-50 of the same book[4], with couplings 0.500, 0.2857 and 0.0774) belongs to this class, and it requires phase shifters with opposite signs.

## 3. PHYSICAL REALIZATION

### 3.1 VTT micron-scale silicon photonics platform

We implemented a 2-stage filter similar to the one in Figure 3 on our micro-scale silicon photonics platform. In fact, after a detailed analysis of constraints and opportunities (see section7), we came to the conclusion that the splitting ratios of that particular design are the most suitable for our platform.

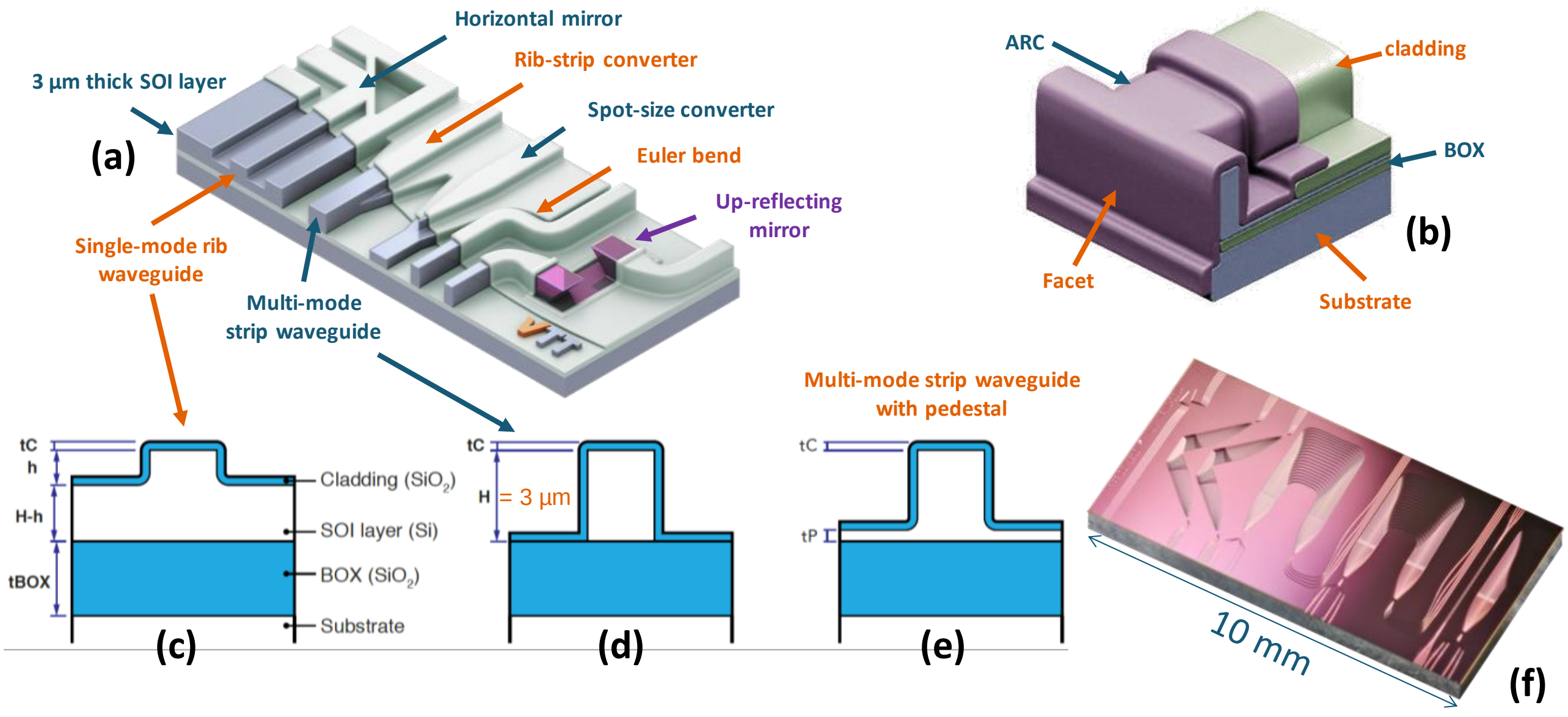

Figure 4. a) Overview of VTT micron-scale silicon photonics platform; b) schematic of the actual T-facet for butt-coupling at the edge of the chip; c), d), e) cross-sections of the main types of available waveguides; f) example of a fabricated chip.

VTT platform is one of the few silicon photonics platforms based on micron-scale thick device layer[14,15], and the only one providing open-access[16], also through multi-project wafer (MPW) runs. We show in Figure 4 an overview of the platform, including the main waveguide types and cross-sections and an example 5×10 $mm^2$ chip with compact arrayed waveguide gratings. The platform has a unique combination of tight bends[17–19] and relatively large modes, enabling dense integration with low propagation losses (≈0.1 dB/cm) and low coupling losses (≈0.5 dB to tapered fibres with 2.5 µm mode field diameter) on a very broad band. Waveguides can be coupled with equally low-loss in-plane, through facets fabricated at wafer scale (Figure 4b), or out-of-plane, through wet-etched 45° up-reflecting mirrors (Figure 4a) based on total internal reflection (TIR). In both cases, an anti-reflection coating is deposited at wafer scale.

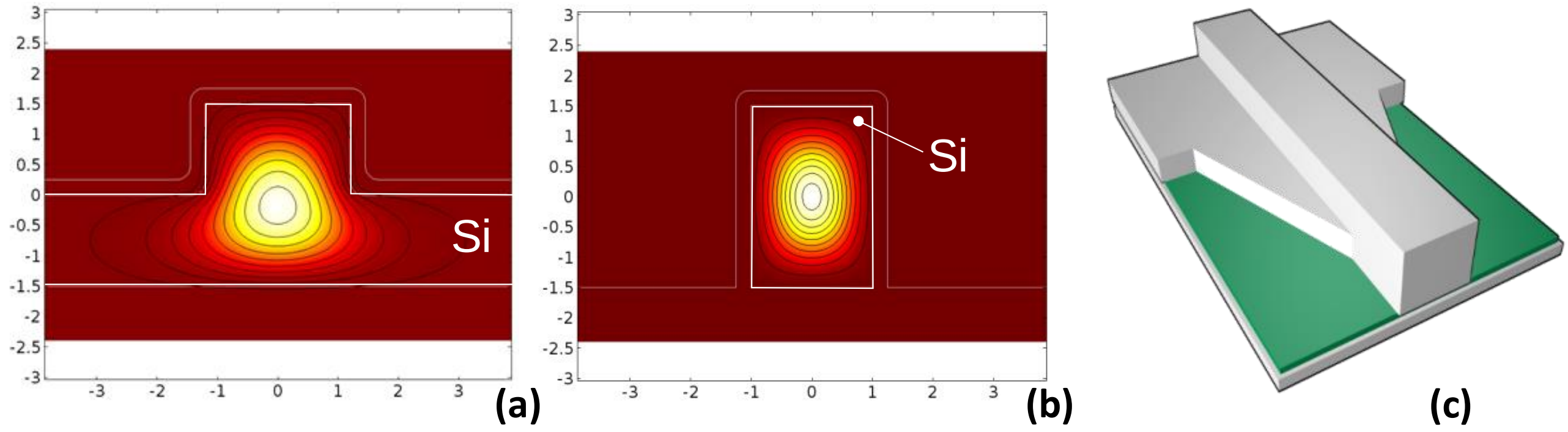

Figure 5. a) Light intensity distribution of the TE mode of a 3 µm thick and 2 µm wide silicon rib waveguide with silica cladding; b) Light intensity distribution of the fundamental TE mode of a 3 µm thick and 2 µm wide silicon strip waveguide with silica cladding; c) 3D sketch of a rib-to-strip converter.

Rib waveguides in silicon layers of any desired thickness can be designed to be singlemode[20] for both polarisations. On the other hand, they are not suitable to achieve tight bends, unless TIR mirrors are used (Figure 4a), with the drawback of non-negligible losses[19] (0.1-0.3 dB/90°). The TE mode of a singlemode rib waveguide is shown in Figure 5a. Instead, strip waveguides are highly multimode, but they allow tight bends. The fundamental TE mode of a strip waveguide is shown in Figure 5b. In order to achieve high integration densities, we tend to prefer strip waveguides in our circuits, even though special care must be taken in order to ensure effective singlemode operation. First of all, input and output couplings happen through singlemode rib waveguides. Then, light is coupled from the rib waveguide to the strip waveguide and vice versa through adiabatic rib-to-strip converters[21], with negligible loss (≈0.02 dB), which excites only the fundamental mode of the strip waveguide. Also the tightest Euler bends[17] are designed to ensure that most of the light remains in the fundamental mode, again resulting in very low-losses (≈0.02 dB).

Another important advantage of the platform, is that all waveguides support TE and TM modes with comparable confinement and losses, meaning that both polarisations are supported and that polarisation insensitive devices can be achieved[14]. Furthermore, the tight confinement of the mode inside silicon, results in reduced sensitivity to fabrication variations[5] compared to submicron waveguides, which is a strong competitive advantage when it comes to actual production yield[15].

### 3.2 The coupling challenge: previous solutions

One of the main challenges in a micron-scale silicon photonic platform is the realisation of compact power splitters. The high mode confinement in the strip waveguide prevents coupling to adjacent waveguides, meaning that directional couplers can be made out of rib-waveguides only. On the other hand, the low lateral index contrast of rib waveguides limits the maximum difference between the eigenmodes of the directional coupler, meaning that directional couplers are typically several millimetres long, also due to the large bending radii in the input and output sections of the coupler. Even more important, etch-depth variations of around ±5% at wafer scale, will result in unacceptable variability of the splitting ratios around the wafer.

For these reasons, we usually resort to strip MMIs that, thanks to the high confinement, operate in our platform very close to the ideal analytical model, also resulting in fairly low losses. MMIs ensure tolerance to fabrication variations and broadband operation, that can be outperformed only by adiabatic couplers[22], but at the expense of significantly larger footprint. Unfortunately, canonical MMIs come with a limited set of splitting ratios, which is a strong limitation when it comes to the design of lattice filters. In the past, we have overcome this limitation by using double MMIs, either with tapered waveguide connections[23] or with bent waveguide connections[8]. As shown in Figure 6, this requires precise phase control both between the two MMIs – so to achieve the wanted splitting ratio – and after the second MMI – so to ensure the $\pi/2$ phase shift between the two arms, like in directional couplers.

The main limitation of this approach is that, even though MMIs are very robust to fabrication variations, in general the two phase shifters are not as robust, resulting in wafer-scale variability of both the splitting ratio and of the output phase. Furthermore, by construction, the double MMIs are more than twice longer than single MMIs and, if the phase shifters are implemented through bent sections (which proved to work better than the tapered waveguides), their layout on mask is complicated by the fact that the output waveguides are in general not collinear with the input waveguides.

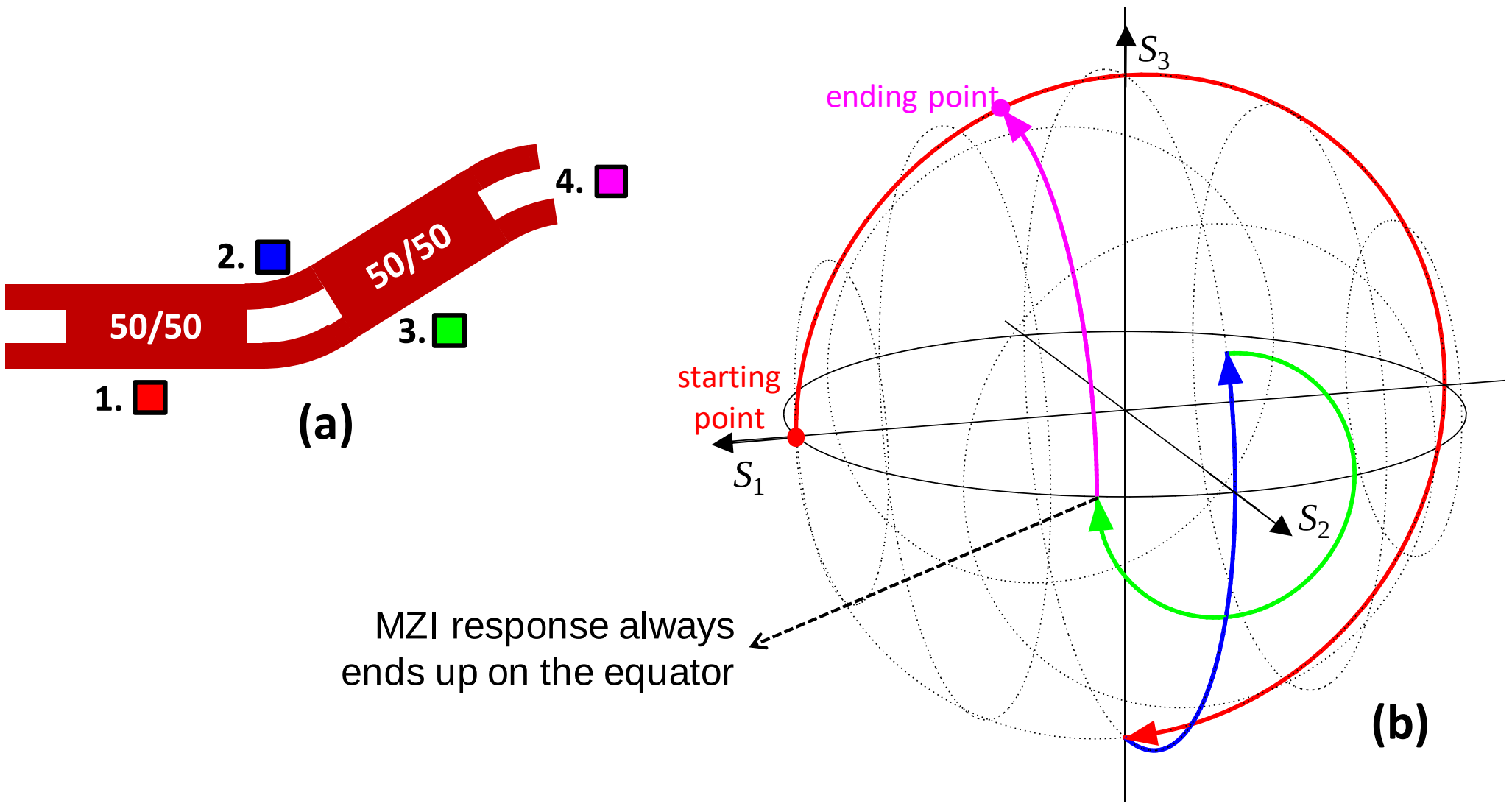

Figure 6. a) Schematic of a double MMI, mimicking the operation of directional coupler; b) geometric representation of its operation principle.

## 3.3 Tapered MMIs

Due to all aforementioned limitations, this time we have decided to resort to a completely different approach, i.e. tapered MMIs. The idea of tapering MMIs to change their splitting ratio has been proposed by Besse et al.[24] back in 1996, and somehow overlooked in silicon photonics, until a systematic experimental study was recently presented by Doménech et al.[25], based on submicron silicon waveguides. By suitably tapering up or down any of the canonical 50:50, 85:15 or 72:28 MMIs, it is possible to achieve any desired splitting ratio with low-loss, broadband operation and high tolerance to fabrication variations. For the first time, we apply this concept to a micron-scale platform, and show how the designed MMIs can achieve less than 0.1 dB loss and a wide operation band. Our goal is to achieve on our platform a robust implementation of a 2-stage generalised lattice filter based on single MMIs. We start noticing that canonical MMIs with other than 50:50 splitting ratios provide 85:15 and 72:28 splitting ratios, where 85 and 72 refer to the cross port. Therefore, also when tapering them, it will be easier to achieve higher coupling to the cross-port, which is at variance with directional couplers. This means that a filter with couplings 0.500, 0.7143, and 0.9226 (Figure 3a) will be easier to implement compared, for example, to a design with couplings 0.500, 0.2857 and 0.0774 (i.e. the other realisation mentioned at the end of section 2.3).

As the 2×2 50:50 MMI, we chose the shortest possible one, that is the one based on general interference[26]. A general rule of thumb to ensure low loss MMIs is to choose input/output waveguides wider than the gap between them, physically meaning that the image does not require too high spatial resolution or, in other words, it does not significantly involve too many high order modes. When considering the resolution limitations of our platform, we ended up with a safe 5 µm wide and 112 µm long realisation, with 1.875 µm wide waveguides and 1.25 µm gaps, as shown in Figure 7a (in order to ease the mask layout of the filter, we have used exactly the same waveguide width and gap for all other types of MMIs). The simulated loss of this design is 0.05 dB, and experimental results show less than 0.2 dB loss. One important thing to keep in mind, is that 50:50 MMIs with general interference come with $-\pi/2$ phase shift at the output, i.e. opposite sign compared to 2×2 50:50 MMIs based on restricted interference[27] or lowest order directional couplers.

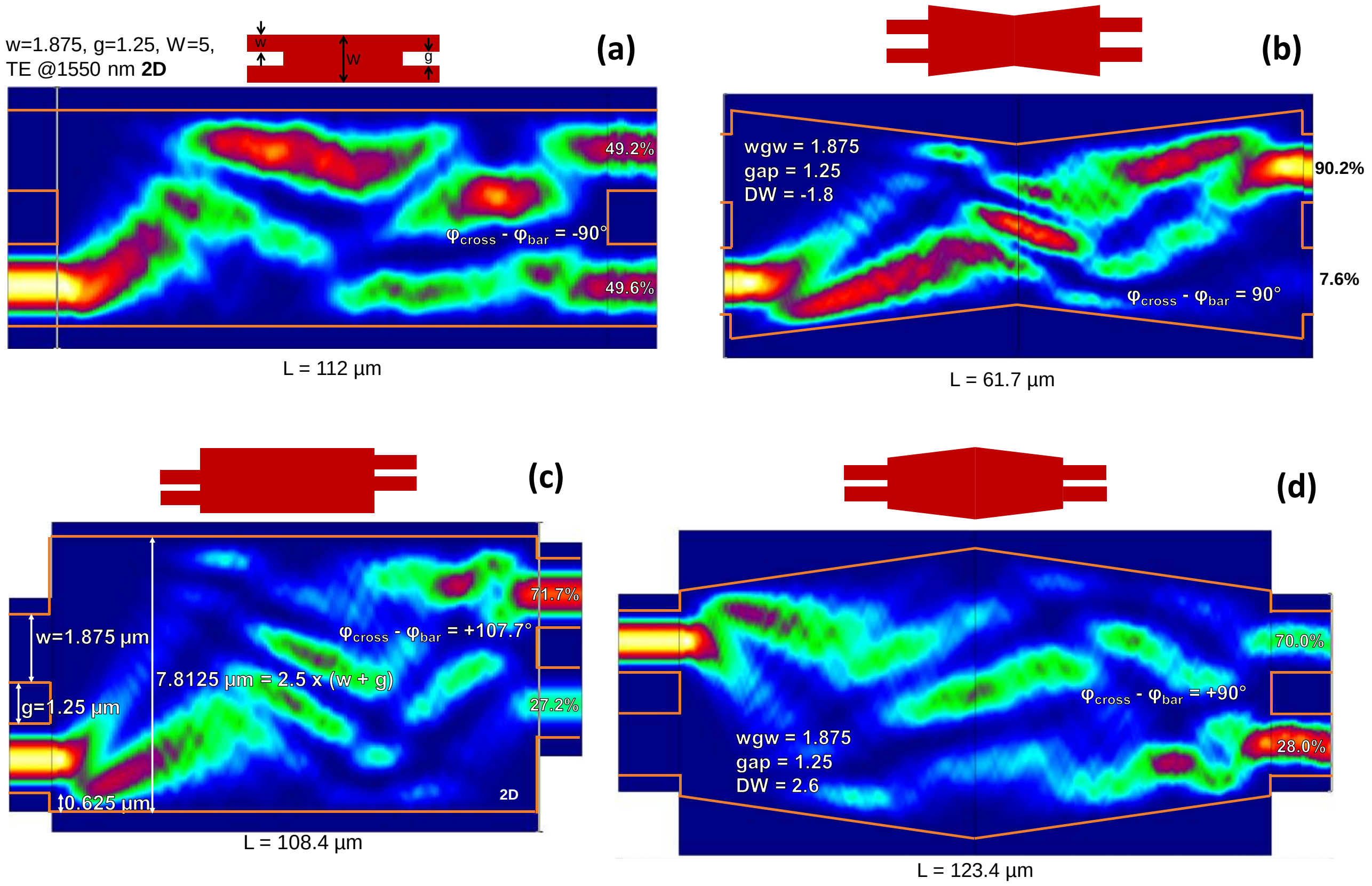

Figure 7. a) The chosen 50:50 MMI with general interference; b) the chosen 92:28 MMI with restricted interference; c) the canonical 72:28 MMI as a possible alternative for the tapered 71:29 MMI depicted in d).

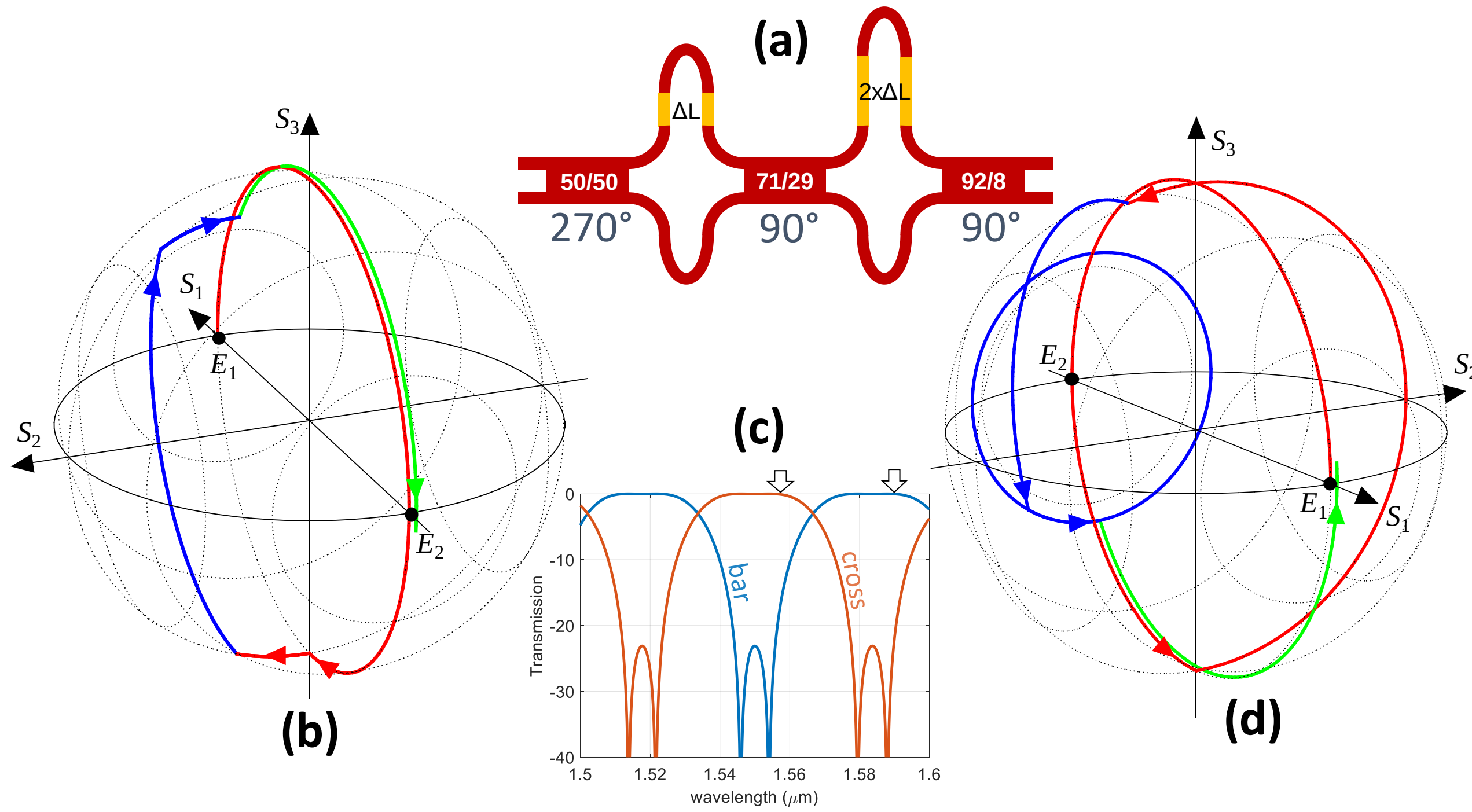

Figure 8. Operation of the updated filter, taking into account the 270° phase at the output of the 50:50 splitter. Otherwise same as in Figure 3.

In terms of the geometric representation, this means that the splitter doesn't reach the north-pole like in Figure 3b and d, but the south pole, instead. When taking this into account, we easily see that the filter must satisfy Eq. (3), instead of Eq. (1) and (2), meaning that, this time, odd and even multiples of $\pi$-phase shifts correspond to light in the bar port and in the cross port respectively. The solution still requires phase shifters with equal sign. The autocorrective operation of the filter on the Bloch sphere is depicted in Figure 8.

Now we need to identify the best choice for the 71:29 splitter. One natural choice would be to resort to the canonical 72:28 splitter shown in Figure 7c. Even though the slightly different splitting ratio wouldn't be a big deal, there are two main reasons why this MMI wouldn't be a good choice. First, the phase at the output is about $3\pi/5$ instead of the desired $\pi/2$, which would require a phase correction element. This would spoil the main justification of the present work, i.e. to avoid any additional phase shifter, so to ensure high tolerance to fabrication variations (as discussed in section 3.2). Another issue with the canonical 72:28 MMI is the offset between the input waveguides and the output waveguides, which poses additional challenges in the mask layout, in order to ensure the right path length of the two phase shifters. This will again require some compensations, which would be again prone to fabrication errors. For this reasons, we resorted instead to the tapered MMI in Figure 7d, which is simply a 85:15 MMI widened by 2.6 µm, resulting in 71:29 splitting ratio with less than 0.1 dB simulated loss. If we instead narrow down the same 85:15 MMI by 1.8 µm, we can achieve 92:8 splitting ratio with less than 0.1 dB simulated loss. The widened MMI is longer than the original 85:15 MMI, whereas the narrowed one is shorter, as it can be easily understood from the scaling law of an MMI length with the square of its width[26]. Both MMIs have the required $\pi/2$ output phase shift, being both based on restricted interference.

### 3.4 Numerical simulations

The MMI simulations in Figure 7 are made with a commercial software (FimmProp by Photon Design) based on the eigenmode expansion method (EME) in 2D, therefore exploiting the effective index method[28]. In practice, we have first calculated the effective index of the fundamental mode of a 3 µm thick silicon slab, and then used it as the refractive index for the silicon region of the 2D simulations. All simulations assume TE operation. The effective index method is typically not very reliable for submicron waveguides, but it is instead very accurate for micron-scale waveguides, with negligible departure from equivalent 3D simulations. This represents a major advantage in terms of calculation time, which is particularly relevant if one wants to simulate the full filter response.

In order to fully exploit the calculation efficiency of EME, we have setup a somewhat artificial simulation, which is physically equivalent to the actual filter operation. First, we have emulated the path difference between the two arms of the phase shifters by choosing a slightly different effective index in the two arms. The change is small enough to avoid any significant back-reflection and high enough to achieve significant phase shifts with a few tens of microns long phase shifter. Second, instead of simulating the wavelength change, we have simply simulated the filter response as a function of the first phase shifter length, the second phase shifter being always double that length. This is physically equivalent to changing the wavelength in the real filter, but it comes with the advantage that the modes of each section are the same for all points of the scan, resulting in a minimal calculation effort. It takes only a few tens of second to calculate the 1000 points of the transmission plot in Figure 9c with a laptop computer. This is a very efficient way to double check if the filter has been designed properly. Indeed, the plot shows the targeted flat-top operation with high extinction ratios.

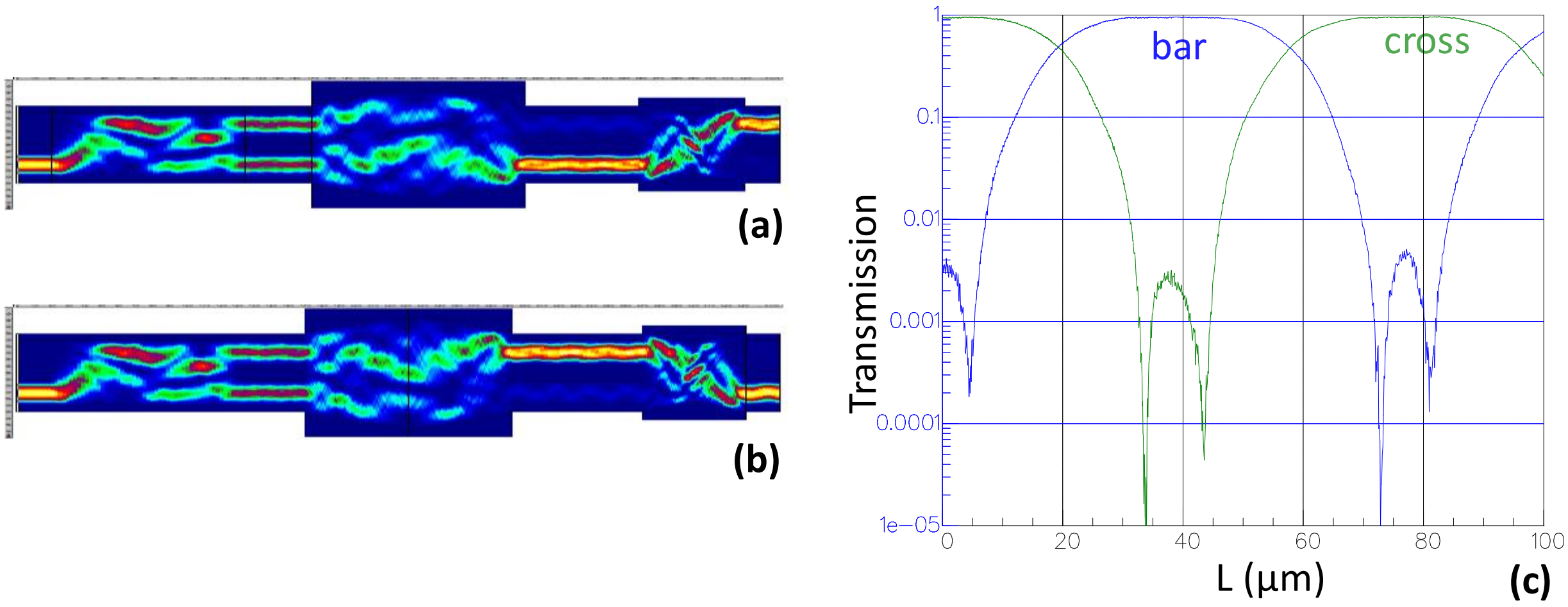

Figure 9. a) simulated 2D map of the emulated filter response for a phase shift close to 2π; b) same for a phase shift close to π; c) filter response as a function of the phase shifter length *L*.

# 4. EXPERIMENTAL RESULTS

## 4.1 Layout of the fabricated devices

After we have designed all the building blocks and verified the correct operation of the designed filter via numerical simulations, we have created a mask layout to fabricate test devices with different FSR and different types of phase shifters. Shorter phase shifters are based on L-bends (central part of Figure 10a and Figure 10b) and longer ones on U-bends (top part of Figure 10a), to ensure smaller footprint. In the bottom part of the chip we have also placed test structures for the single MMIs.

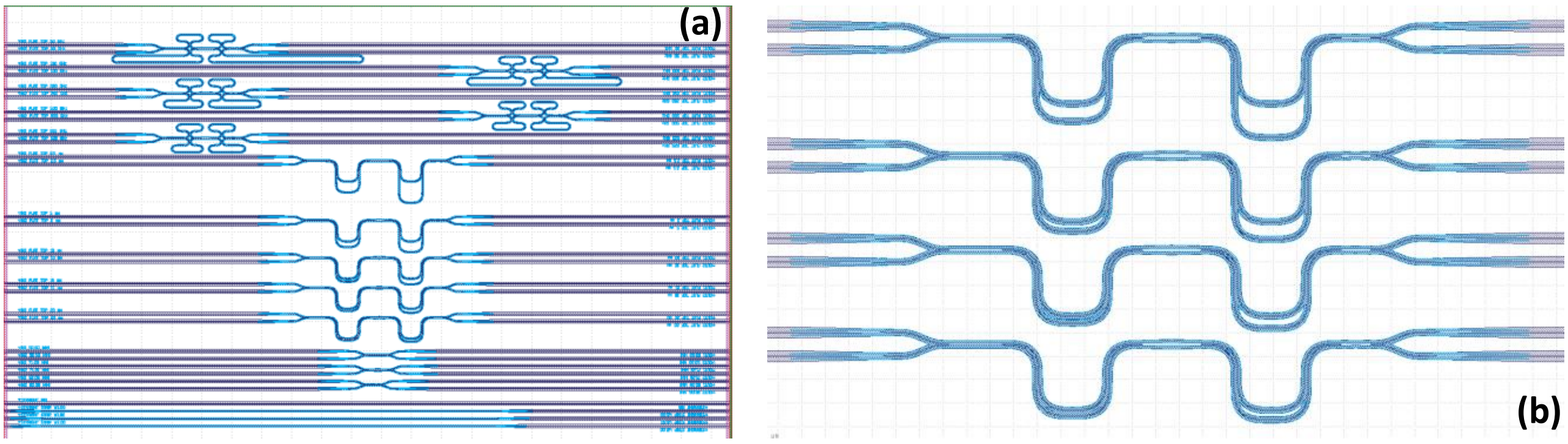

Figure 10. a) Mask layout of different implementations of the flat top filters and their building blocks and b) Close-up of the filters based on L-bends.

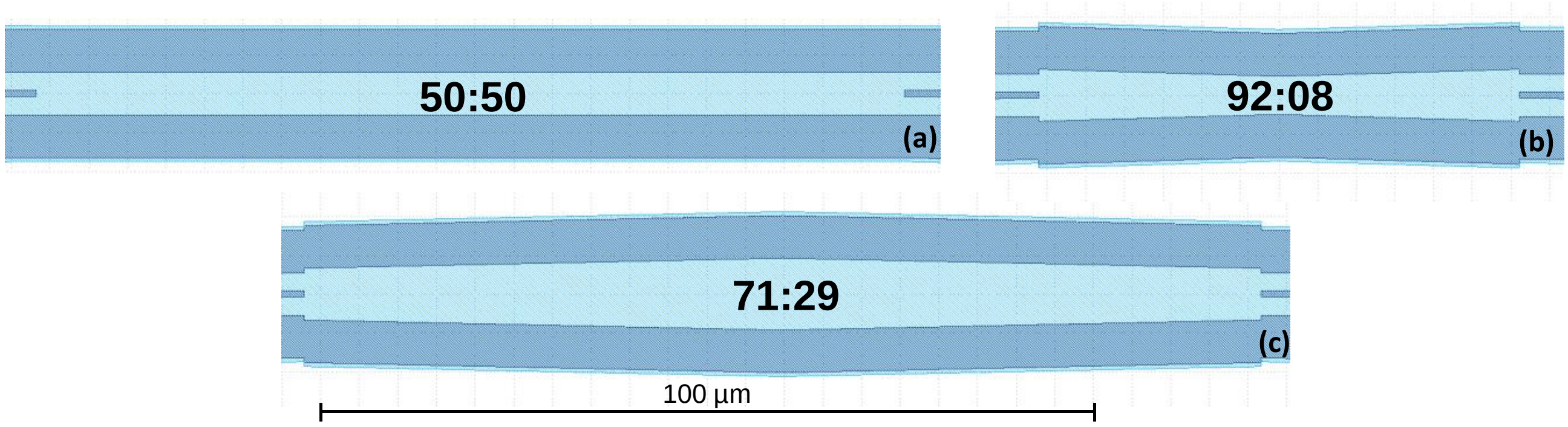

Figure 11. Close-up of the three MMI layouts. They are all in scale.

Input and output waveguides are singlemode rib waveguides, whereas the filter is all based on through etched strip waveguides with a 200 nm pedestal (Figure 4e). We show in Figure 11 also a close-up of the three MMIs we used in all different filters.

### 4.2 Fabrication in a MPW run

The devices were fabricated on our platform as part of a MPW run. We used smart-cut silicon-on-insulator wafers from SOITEC. Initial SOI layer thickness was increased with epitaxial silicon growth to about 3.2 µm. FilmTek 4000 spectrophotometer was used to measure the thicknesses of the SOI layer and the underlying buried oxide. A 500 nm thick Tetraethyl orthosilicate (TEOS) layer was deposited on the wafer in LPCVD diffusion furnace to work as a hard mask in silicon etching. Waveguide fabrication was done using our standard double-masking multi-step process[29]. In the process, two mask layers are passively aligned with respect to each other, and two separate silicon etch-steps form the rib and strip waveguide structures and waveguide facets. Lithography steps were done using FPA-2500i4 i-line wafer stepper from Canon Inc. and pattern transfer to oxide hard mask was done using LAM 4520 reactive ion etcher with $CF_4$ and $CHF_3$ chemistry. Waveguides were etched into silicon using Omega i2l ICP etcher from SPTS Technologies. The etching was done with a modified Bosch process[30] using $SF_6$ and $C_4F_8$ as etch and passivation gases, respectively, and $O_2$ as an etch gas to break passivation polymer formed by $C_4F_8$. After silicon etching, TEOS hard mask was removed with buffered oxide etch (BOE). Hard mask removal was followed by wet thermal oxidation consuming 225 nm of silicon, and thermal oxide removal with BOE. This was done to smoothen the etched surfaces, and to thin the SOI-layer to its final thickness of approximately 3 µm. A 190 nm thick silicon nitride layer was then deposited on the wafer with LPCVD as an antireflection coating and to prevent the etching of buried oxide in later oxide wet etch process steps. This layer was patterned in hot phosphoric acid using a hard mask made of 250 nm thick LPCVD TEOS patterned with BOE. After silicon nitride layer patterning, 265 nm of LPCVD TEOS was deposited as a cladding layer for the waveguides. As a last step, the oxide layers were removed from the waveguide facets with BOE etch, and wafer was diced into chips.

### 4.3 Optical characterisation and analysis of the results

We have characterised the spectral response of the MMIs and of the filters with a tunable laser centred around 1550 nm wavelength. The MMIs show robust operation in the target wavelength range, as clearly shown in Figure 12. Remarkably, they work pretty well also for TM polarisation, even though they had been designed for TE operation only. The measured spectral response of a full filter with 30 nm FSR is shown in Figure 13, showing flat-top behaviour for both TE and TM polarisation. The transmission is normalised to the one of a 5 mm long straight rib waveguide. The losses are significantly higher than expected, and extinction ratios are lower than simulated.

We have identified two major problems that limited the filter performance. A first issue is the bend losses when using strip waveguides with a pedestal, like the one depicted in Figure 4e. First, in order to allow for about ±150 nm etch-depth variations at wafer scale, the actual thickness of the pedestal is most of the time thicker than the nominal 200 nm value. Second, the thermal oxidation step (see previous section) induces a significant rounding of the corners (also sketched in Figure 4e). Even though simulations of an ideal strip waveguide bend with nominal 200 nm pedestal and no corner rounding show negligible losses, when taking into account thickness variations and corner rounding, we find that losses are not negligible, and especially high for tighter bends. Even though we have now changed our technology and completely removed the pedestal from all strip waveguides that don't require lateral implantations, unfortunately we fabricated these filters in a run still based on the old technology. That very same run has been particularly unlucky also in terms of poor linewidth control, which caused the second major issue: all MMIs have been fabricated about 200 nm wider than designed, which moved the nominal operation wavelength to about 1600 nm. Indeed, the spectra in Figure 13 show a clear trend of lower losses towards longer wavelengths.

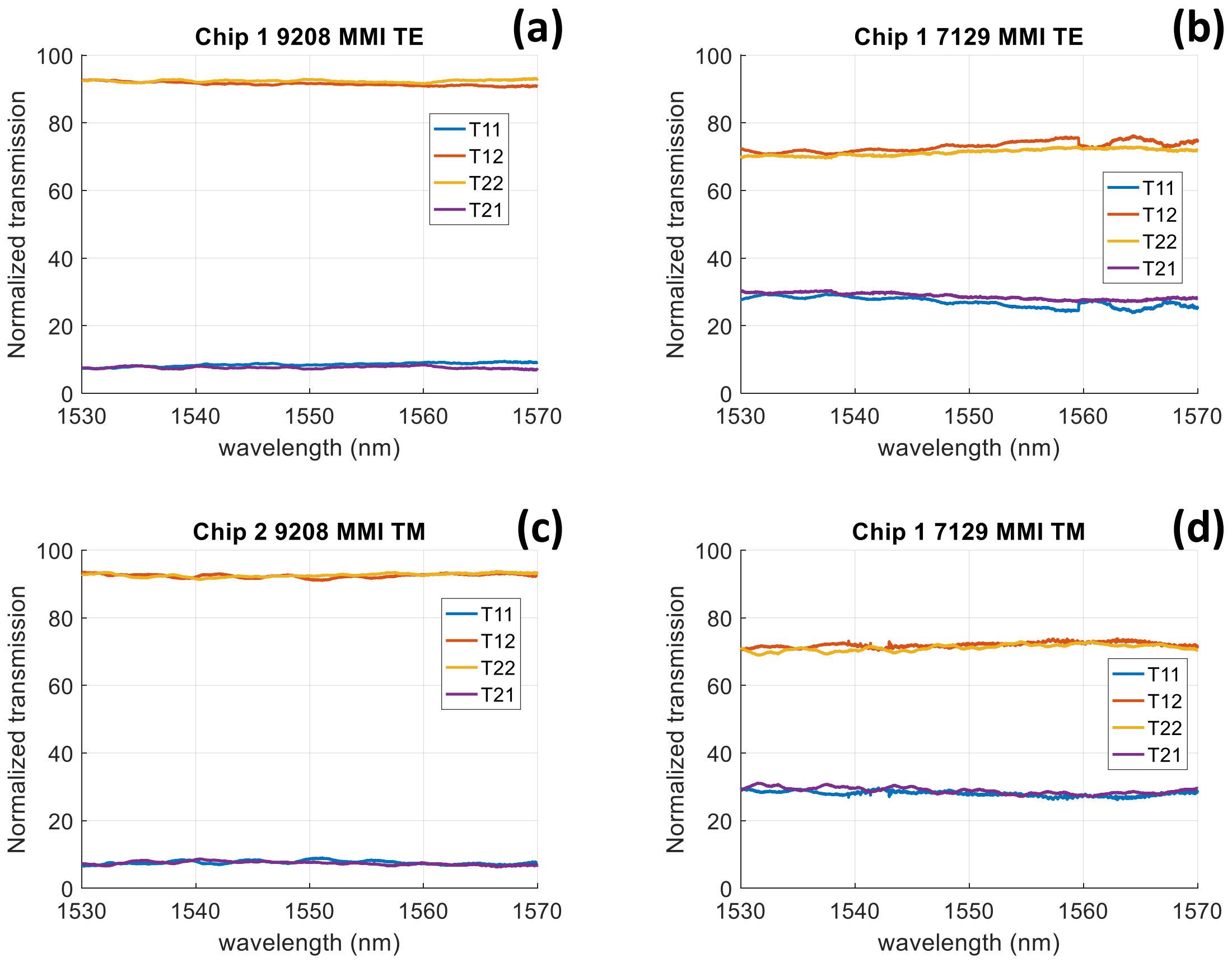

Figure 12. Characterisation of the tapered MMIs over a 40 nm range around 1550 nm, for both TE and TM polarisation.

Nevertheless, we stress three remarkable facts. First, the spectra still resemble the nominal design, despite all the aforementioned issues. Second, the filters work also for the other polarisation, even though they have been designed for TE operation only. Third, we stress that no heater was used to align the operation of the two phase shifters, that worked just as designed. All these evidences are a clear proof of the robustness of the proposed design.

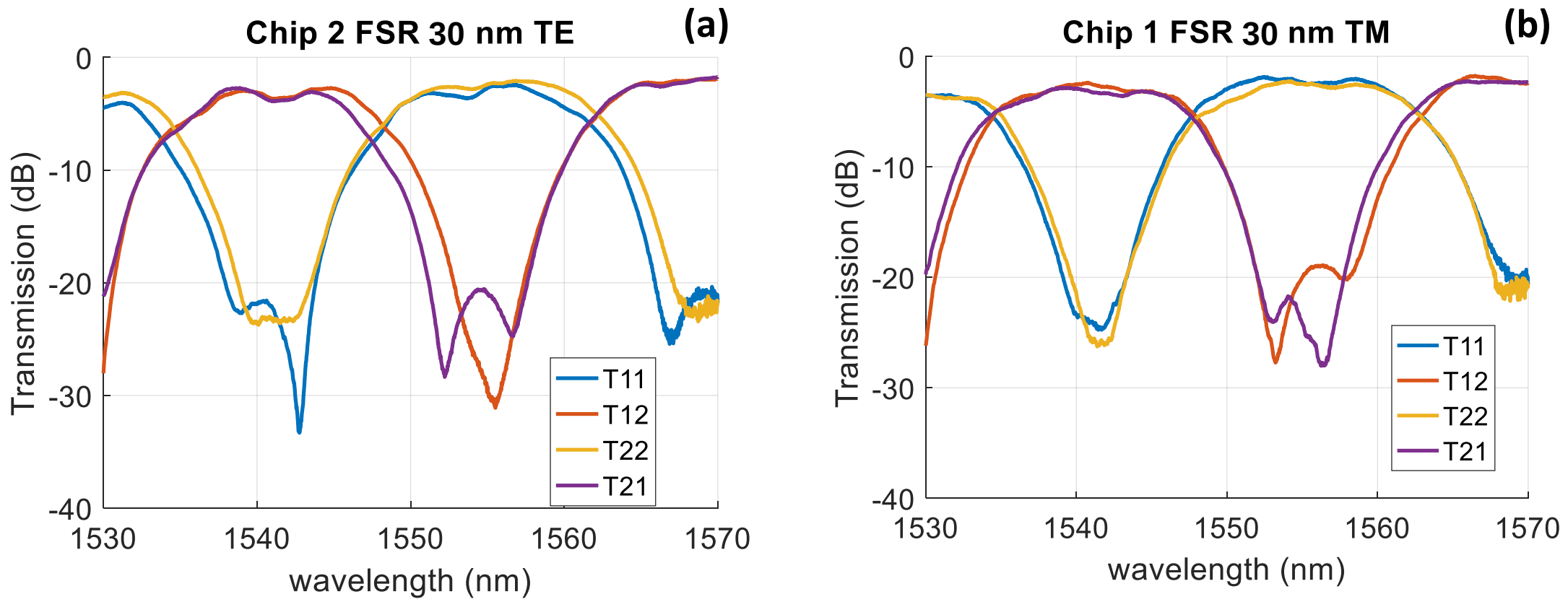

Figure 13. Characterisation of an interleaver with 30 nm FSR for both TE and TM polarisation.

# 5. CONCLUSION

We have analysed flat-top filters based on 2-stage generalised lattice filters in terms of a geometrical representation, gaining significant physical insight on their working principle. This enabled us also to derive simple analytical formula for the design of such filters. Thanks to the gained physical insight, we have been able to identify a filter design that is particularly suitable to be implemented on our micron-scale platform based on single MMIs. We have analysed different possible alternatives for the MMIs to match the desired splitting ratios, and eventually opted for compact and low-loss tapered MMIs. We confirmed, by efficient numerical simulations, that 2-stage filters based on the designed MMIs operate properly as flat-top filters. We then drew variations of the designed filter on a mask layout, for them to be fabricated in a MPW run. Optical characterisation of the fabricated devices demonstrated robust broadband flat-top operation of both the MMIs and the filters, despite the encountered fabrication issues. We have therefore successfully identified for the first time a robust approach to design and fabricate flat-top filters based on single MMIs.